\documentclass[runningheads]{svmult}

\usepackage{makeidx}   
\usepackage{graphicx}  
\usepackage{subeqnar}  
\usepackage{multicol}  
\usepackage{physprbb}  
\makeindex             

\def\om{\Omega_p}

\def\kin{{\cal K}}
\def\pin{{\cal P}}
\def\smt{{\cal S}}
\def\DV{{\Delta V}}

\def\spose#1{\hbox to 0pt{#1\hss}}
\def\gtsim{\mathrel{\spose{\lower.5ex \hbox{$\mathchar"218$}}
     \raise.4ex\hbox{$\mathchar"13E$}}}
\def\ltsim{\mathrel{\spose{\lower.5ex\hbox{$\mathchar"218$}}
     \raise.4ex\hbox{$\mathchar"13C$}}}

\def\degrees{^\circ}

\def\kms{$\mathrm km\;s^{-1}$}
\def\kmsk{$\mathrm {km}~\mathrm{s}^{-1}~ \mathrm{kpc}^{-1}$}

\begin{document}
\title*{A Pattern Speed in the Galaxy's OH/IR Stars}
\toctitle{A Pattern Speed in the Galaxy's OH/IR Stars}
%
%
\titlerunning{A Pattern Speed in the Galaxy's OH/IR Stars}
%
\author{Victor P. Debattista\inst{1}
\and Ortwin Gerhard\inst{1}
\and Maartje N. Sevenster\inst{2}}
\authorrunning{Victor P. Debattista et al.}
%
%
\institute{Astronomisches Institut, 
     Universit\"at Basel, 
     Venusstrasse 7,
     CH-4102 Binningen, Switzerland
\and RSAA/MAAAO, RSAA/MSSSO, 
     Cotter Road, 
     Weston ACT 2611, Australia}

\maketitle              


\section{Introduction}
The Milky Way Galaxy (MWG) contains both a bar and spirals.  The pattern
speed/rotation frequency, $\om$, of these components have been measured with
a variety of models.  For the bar, models have found values 
$40 \ltsim \om \ltsim 60$ \kmsk\ (Binney et al. 1991; Fux 1999; 
Englmaier \& Gerhard 1999; Weiner \& Sellwood 1999; Dehnen 1999; Bissantz
et al. 2002).  The spiral arm $\om$ is even more uncertain, with 
values in the range $13.5 \ltsim \om \ltsim 27$ \kmsk\ reported (Morgan 1990; 
Amaral \& L\'epine 1997; Mishurov \& Zenina 1999).

A model-independent method, based on the continuity equation, for measuring 
pattern speeds in external galaxies has been developed by Tremaine \& 
Weinberg (1984).  This method can be extended to the MWG (Kuijken \& 
Tremaine 1991; Debattista et al. 2002).  For discrete tracers in the MWG, 
the Tremaine-Weinberg (TW) method is contained in the following expression:
\begin{eqnarray}  
\DV & \equiv & \om R_0 - V_{\rm LSR} \equiv \ 
    \frac{\kin}{\pin} - u_{\rm LSR} \frac{\smt}{\pin} \nonumber \\
    & = & \frac{ \sum_{i} f(r_i)v_{r,i}}{\sum_{i} f(r_i)\sin l_i 
\cos b_i} - u_{LSR} 
\frac{ \sum_{i} f(r_i)\cos l_i \cos b_i}{\sum_{i} f(r_i)\sin l_i \cos b_i} 
\label{eqn1}
\end{eqnarray}
where $R_0$ is the Sun-MWG center distance, $V_{\rm LSR}$ is the tangential
velocity of the local standard of rest (LSR), $u_{\rm LSR}$ is the radial
velocity of the LSR, $f(r_i)$ is the observational detection probability 
(which need not be known), $v_{r}$ is the heliocentric radial velocity of 
a discrete tracer, and $(l,b)$ are its Galactic coordinates.  Eqn. \ref{eqn1} 
assumes only one pattern speed and a low amplitude for any rapidly
growing structure.  Moreover, the tracer population needs to be sampled 
uniformly; one such survey is the ATCA/VLA OH 1612 MHz 
survey (Sevenster et al. 1997a,b \& 2001), covering 
$|l| \leq 45\degrees$ and $|b| \leq 3\degrees$.

\section{Pattern Speed of the OH/IR Population}

\begin{figure}[ht]
\begin{center}
\includegraphics[width=.2\textwidth,angle=-90]{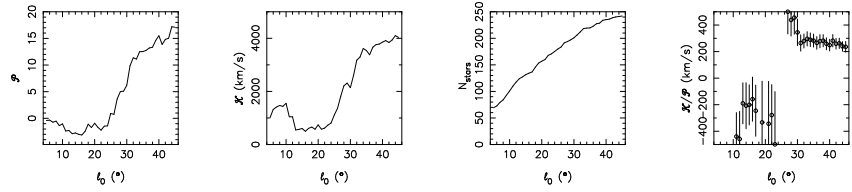}
\end{center}
\caption[]{The TW analysis for the OH/IR stars for changing $l_0$ (the
maximum $|l|$ in Eqn. \ref{eqn1}).  From left to right are
$\pin$, $\kin$, the number of stars and the resulting $\kin/\pin$}
\label{fig1}
\end{figure}

We extracted from the ATCA/VLA OH 1612 MHz survey a sample of $\sim 250$ 
OH/IR stars which are relatively old (older than 
0.8 Gyr) and bright (flux density greater than 0.16 Jy).  These selection 
criteria give OH/IR stars between 4 and 10 kpc away from the Sun.  We applied 
the TW analysis of Eqn. \ref{eqn1} to this sample, obtaining the
results shown in Fig. \ref{fig1} (Debattista et al. 2002).  Note that the 
value of $\kin/\pin$ has
converged, within the errors, for $|l| > 30\degrees$.
Simple tests with models show that a measurement of $\DV$ with a sample of 
this size should give an average accuracy of $\sim 17\%$ and always better 
than $40\%$, when the asymmetry signal is as large as the one we find.  
From re-sampling experiments, we find $\DV = 252 \pm 41$ \kms.  If
we assume $V_{\rm LSR}/R_0 = 220/8$ \kmsk\ (from SgrA$^*$ motion, 
Backer et al. 1999; Reid et al. 1999 and Cepheid proper motions, Feast
\& Whitelock 1997) and $u_{\rm LSR} = 0$ (from SgrA$^*$ 
HI absorption spectrum, Radhakrishnan et al. 1980), we obtain 
$\om = 59 \pm 5$ \kmsk.  We estimate systematic error to be $\sim 10$ \kmsk.

The signal we found is concentrated close the plane ($|b| \leq 1\degrees$)
and at large longitude, suggesting a spiral is responsible for it (possibly
the Scutum arm).  The high $\om$ suggests this spiral arm is driven by the 
bar.  Alternatively, the non-axisymmetric structure involved 
is an inner ring (Sevenster \& Kalnajs 2001).

%

\end{document}